\documentclass[journal]{IEEEtran}
\usepackage{mathrsfs}
\usepackage{bbm}
\usepackage{amssymb}
\usepackage{bbding}
\usepackage{threeparttable}
\usepackage{diagbox}

\usepackage[mathcal]{euscript}

\usepackage{psfrag,calc,url,bm}

\usepackage{cite}

\usepackage{graphicx}

\usepackage{psfrag}

\usepackage{subfigure}
\usepackage{hyperref}
\usepackage{url}

\usepackage{stfloats}

\usepackage{amsmath}

\usepackage{float}

\usepackage{algorithmic}

\usepackage[ruled,linesnumbered,vlined]{algorithm2e}

\usepackage{color}

\usepackage{boxedminipage}

\usepackage{amsthm}

\usepackage{multirow}

\usepackage{setspace}

\usepackage{soul}
\usepackage{epstopdf}

\usepackage{xcolor}
\usepackage{tikz,pgfplots,filecontents}

\allowdisplaybreaks[3]

\begin{document}

\title{Model-Based GNN Enabled Energy-Efficient Beamforming for Ultra-Dense Wireless Networks}
\author{Rongsheng Zhang, Yang Lu,~\IEEEmembership{Member,~IEEE}, Wei Chen,~\IEEEmembership{Senior Member,~IEEE}, \\Bo Ai,~\IEEEmembership{Fellow,~IEEE}, and Zhiguo Ding,~\IEEEmembership{Fellow,~IEEE}
\thanks{Rongsheng Zhang and Yang Lu are with the School of Computer Science and Technology, Beijing Jiaotong University, Beijing 100044, China (e-mail: 20271054@bjtu.edu.cn, yanglu@bjtu.edu.cn).}
\thanks{Wei Chen and Bo Ai are with the School of Electronic and Information Engineering, Beijing Jiaotong University, Beijing 100044, China (e-mail: weich@bjtu.edu.cn, boai@bjtu.edu.cn).}
\thanks{Zhiguo Ding is with Department of Electrical Engineering and Computer Science, Khalifa University, Abu Dhabi 127788, UAE (e-mail: zhiguo.ding@ieee.org).}
}

\maketitle


\begin{abstract}
This paper proposes a novel deep learning enabled beamforming design for ultra-dense wireless networks by integrating prior knowledge and graph neural network (GNN), termed model-based GNN. An energy efficiency (EE) maximization problem is first subject to the budget and quality of service (QoS) requirements, and then reformulated based on the minimum mean square error  scheme and the hybrid zero-forcing and maximum ratio transmission schemes. Based on the reformulated problem, a model-based GNN is designed to realize the mapping from channel state information to beamforming vectors. Particular, the multi-head attention mechanism and residual connection are adopted to enhance the feature extracting, and a scheme selection module is designed to improve the adaptability of GNN. The unsupervised learning is adopted, and a various-input training strategy is proposed to enhance the stability of GNN. Numerical results demonstrate that the proposed GNN scheme can realize a millisecond-level response with limited performance loss, the scalability to different users and the adaptability to various channel conditions and QoS requirements of the model-based GNN in ultra-dense wireless networks.

\end{abstract}


\section{Introduction}

\subsection{Background}

With the rapid development of the mobile intelligent applications, an increasing number of studies on the sixth-generation (6G) technology, predominantly from industry and academia, are focusing on how to enhance the wireless converge for ultra-dense wireless networks, e.g., space-air-ground-sea integrated networks\cite{sags} and Internet of Everything \cite{IoE}. The time-intensive and high quality of service (QoS) requirements of the mobile intelligent applications further the services of the wireless communication systems. However, it is challenging to intelligent resource allocation for ultra-dense wireless networks due to a huge amount of variables required to be optimized \cite{intro1}. Heavy communication overheads and huge computational complexities hinder the implementation of the traditional algorithms such as convex optimization based (CVX-based) approaches for small-scale optimization. Besides, the time-varying and dynamic nature of wireless networks urgently requires more intelligence in the resource allocation algorithms to enhance the service capability flexible to various scenarios with diverse requirements \cite{cv-channel,new1}. 

Recently, deep learning (DL) has shown to achieve superb behavior in wireless resource allocation following the learning-to-optimize paradigm due to its superior learning capability and high computational efficiency \cite{AI}. Particularly, the DL-enabled transmission design was shown able to achieve near-optimal performance and real-implementation in interference channel based on multi-layer perceptron (MLP) \cite{MLP1} and millimeter wave  multiple-input-multiple-output (MIMO) systems based on convolutional neural network (CNN) \cite{CNN1}. However, the generalization capability of MLP and CNN is limited due to their fixed input dimensions, especially for the scenarios unseen in the training set. Therefore, some existing attempted to develop specialized neural networks for wireless networks based on graph neural network (GNN) since the wireless networks are graph-topology \cite{lugnn}. Nevertheless, for the beamforming design in ultra-dense wireless networks, plenty of optimization variables are needed to construct more complex GNN architectures, which complicates the training process and degrades the inference speed. Instead of data-driven DL, the model-driven DL allows to leverage prior knowledge to simplify the mapping constructed by neural networks to realize a balance trade-off between light weighting  and capacity of neural networks \cite{zhangtvt}. Therefore, the model-based GNN should be an attractive solution approach for the resource allocation problem in ultra-dense wireless networks. 

In addition, green communication has been the central consideration in the ultra-dense wireless networks since the fifth-generation (5G) era \cite{green}. Consequently, energy efficiency (EE) is regarded as a crucial performance metric in wireless resource allocation \cite{EE-key}. Some related works have reveal the trade-off between spectral efficiency and EE. By adding the QoS requirements, such a trade-off can be balanced \cite{SE-EE}. Besides, the energy efficient design can prolong the operation lifetime of the ultra-dense wireless networks. Thus, jointly enhancing the service capability and EE of ultra-dense wireless networks becomes a central goal for DL-enabled transmission, which will be focused on in this paper. 

\subsection{Related works}

Thus far, the spectral or energy efficient wireless resource allocation algorithms have been widely investigated while most of them are from the perspective of the conventional CVX-based approach. For example, in \cite{SOCP}, an accelerated projected gradient (APG) algorithm based on the second-order cone programs (SOCPs) was proposed to maximize the total EE under in cell-free massive MIMO. In \cite{Dinkelbach}, the Dinkelbach method was utilized to maximize the EE for multi-user MIMO systems with dynamic metasurface antennas. In \cite{SDP}, the energy Efficient beamforming and cooperative jamming   algorithm was proposed based on semi-definite programming (SDP) for the perfect channel state information (CSI) case and S-procedure for the imperfect CSI. In \cite{SCA} and \cite{BSUM}, the successive convex approximation (SCA) and block successive upper bound minimization (BSUM) were respectively proposed for multiple-input-single-output (MISO) interference channels to maximize  the sum rate with the constraints of the tolerable transmission outage. However, the iterative framework and the memoryless nature of CVX-based approaches hinder the implementation of these CVX-based approaches in time-varying wireless environment.     





Therefore, some recent works attempted to develop GNN-enabled transmission designs. In \cite{WCGCN}, a wireless channel graph convolution network (WCGCN) based on message passing graph neural network (MPGNN) was proposed to solve the large-scale radio resource management problem for weighted sum-rate maximization. The permutation equivalence of WCGCN and the equivalence between MPGNN and WMMSE were analyzed. Besides, a bipartite GNN (BGNN) \cite{BGNN} was proposed to facilitate scalable hybrid beamforming in multi-user MISO (MU-MISO) networks, and a heterogeneous graph neural network (HGNN) \cite{HGNN} was proposed to tackle the spectral efficiency maximization problem for a multi-carrier-division cell-free massive MIMO. However, the feature extracting of this existing work was based on the vanilla message passing. To enhance the feature extracting and mitigate the over-smoothing issues, some works attempt to design more sophisticated GNNs to realize end-to-end beamforming design. In \cite{dl,dl2,dl3}, the multi-head attention mechanism and residual connection were integrated into message passing to solve the beamforming design with the aims of EE maximization, sum-rate maximization and max-min rate, respectively. In \cite{icnet}, a graph attention neural network based model was proposed for energy efficient beamforming design in the interference channels. In \cite{new2}, an HGNN with graph attention and semantic attention was designed realize physical-layer secure beamforming. The effectiveness of  multi-head attention mechanism and residual connection was numerically validated in various scenarios. Nevertheless, the complicated GNNs may degrade the computational efficiency which limites the implementation of these end-to-end GNNs in ultra-dense wireless networks. In a summary, both existing CVX-based and GNN-based approaches may not work in ultra-dense wireless networks.   


\subsection{Contributions}



This paper aims to develop a model-based GNN enabled energy-efficient beamforming design for ultra-dense wireless networks to realize near-optimal and real-time optimization with excellent scalability to  users and adaptability various channel conditions and QoS requirements. An EE maximization problem is formulated and solved by the proposed model-based GNN which includes three ingredients, i.e., problem reformulation with prior knowledge, architecture of GNN and learning framework. The main contributions are summarized as follows: 
\begin{enumerate}
    \item The EE maximization problem is reformulated based on following two schemes, namely, the minimum mean square error (MMSE) scheme and the hybrid zero-forcing  and maximum ratio transmission (HZM)  schemes \cite{hyb}, to reduce the number of optimization variables with limited performance loss. 
    \item Based on the reformulated problems, we propose a model-based GNN to realize the mapping from CSI to beamforming vectors. Particular, the multi-head attention mechanism and residual connection are adopted to enhance the feature capability of the designed GNN. Furthermore, the scheme selection module is designed to improve the adaptability of GNN. 
    \item The unsupervised learning is adopted to train the model-based GNN. The power budget constraint is guaranteed via activation functions while the QoS requirements are satisfied via penalty method. To enhance the stability of GNN in various application scenario, a various-input training strategy is proposed.
\end{enumerate}

Extensive simulation results are provided to demonstrate
the efficacy of the proposed model-based GNN. In particular, the key properties of the proposed GNN scheme, including millisecond-level response with limited performance loss, the scalability to different users and the adaptability to various channel conditions and QoS requirements, are numerically validated for the proposed model-based GNN. The MMSE and HZM schemes are compared and analyzed. Besides, the various-input training strategy is shown to improve the stability of the GNN.  

The rest of this paper is organized as follows. Section II introduces the system model and problem formulation. The model-based GNN is presented with problem reformulation based on MMSE and HZM schemes, architecture of GNN and learning framework in Section III. Simulation results are given in Sections IV. Finally, Section V concludes this paper.

\emph{Notations}: The following mathematical notations and symbols are used throughout this paper. $\bf a$ and $\bf A$ stand for a column vector and a matrix (including multiple-dimensional tensor), respectively. The sets of real numbers and n-by-m real matrices are denoted by ${\mathbb{R}}$ and ${\mathbb{R}^{n \times m}}$, respectively. The sets of complex numbers and $n$-by-$m$ complex matrices are denoted by ${\mathbb{C}}$ and ${\mathbb{C}^{n \times m}}$, respectively. For a complex number $a$, $\left| a \right|$ denotes its modulus. ${{\rm Re}(a)}$  and  ${{\rm Im}(a)}$  denote its real and imaginary part, respectively. For a vector $\bf a$, ${\left\| \bf a \right\|}$ is the Euclidean norm. For a matrix ${\bf A}$, ${\bf A}^H$  denotes the conjugate transpose, respectively.

\section{System Model and Problem Formulation}

\begin{figure}[t]
\begin{center}
{\includegraphics[ width=.5\textwidth]{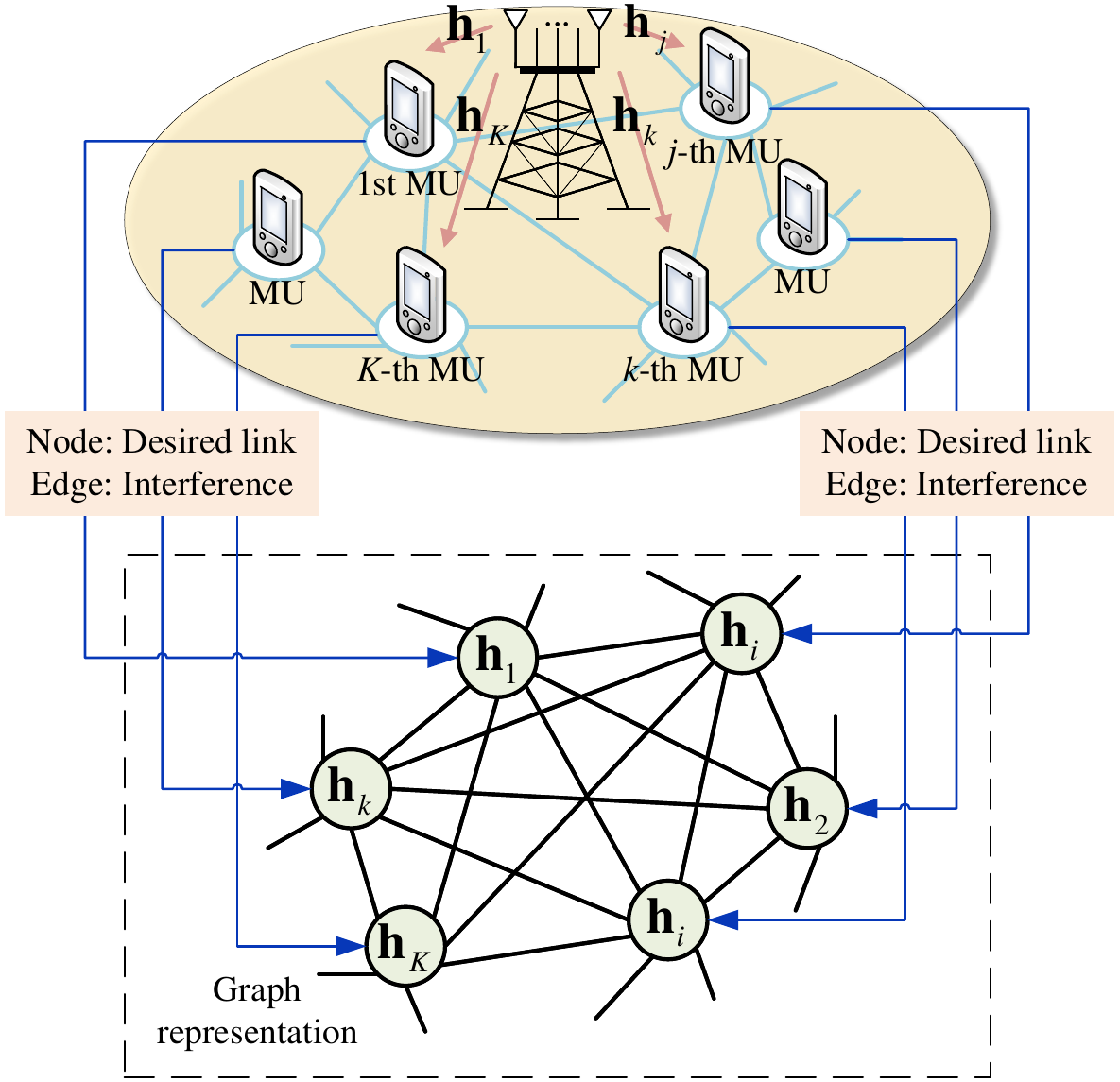}}
\caption{A graph representation of the MU-MISO system.}
\label{sys}
\end{center}
\end{figure}
Consider a downlink MU-MISO network, as shown in Figure \ref{sys}, where one $N_{\rm T}$-antenna transmitter serves $K$ single-antenna mobile users (MUs) over a common spectral band. We use $\mathcal{K} \triangleq \{1,2, ..., K\}$ to denote the index set of the MUs. The transmit signal from the transmitter is given by
\begin{flalign}
{\bf x} =  \sum\nolimits_{i \in \mathcal{K}} {{{\bf{w}}_i}{s_i}},
\end{flalign}
where $s_k$ and ${{{\bf{w}}_k}}$ denote the symbols for the $k$-th MU and the corresponding beamforming vector, respectively. Without loss of generality, it is assumed that ${{\mathbb E}}\{ {{{| {s_i } |}^2}} \} = 1$ ($\forall i\in\mathcal{K}$). For transmitting $\bf x$, the total power required by the transmitter is given by 
\begin{flalign}\label{power}
P_{\rm Total}\left( \{{{\bf{w}}_i}\} \right) = { \sum\nolimits_{i \in \mathcal{K}} {\left\| {{{\bf{w}}_i}} \right\|^2}  + {P_{\rm C}}},
\end{flalign}
where ${P_{\rm C}}$ denotes the constant power consumption introduced by circuit modules.

Denote the CSI of the $k$-th transmitter-MU link by ${{{\bf{h}}_k}}\in{\mathbb{C}^{{N_{\rm{T}}}}}$. The received signal at the $k$-th MU is given by
\begin{flalign}
\label{signal}
{{\rm{y}}_k} =  {\bf{h}}_k^H{\bf x} ={\bf{h}}_k^H{{\bf{w}}_k}{s_k} + \sum\nolimits_{i \in \mathcal{K}/\{k\}} {{\bf{h}}_k^H{{\bf{w}}_i}{s_i}}  + {n_k},
\end{flalign}
where $n_k\sim\mathcal{CN}( {0,{\sigma_k ^2}})$ denotes the additive white Gaussian noise (AWGN) at the $k$-th MU.  Then, the achievable rate at the $k$-th MU is expressed as
\begin{flalign}\label{rate}
{R_k}\left( {\left\{ {{{\bf{w}}_i}} \right\}} \right) = {\log _2}\left( {1 + \frac{{{{\left| {{\bf{h}}_k^H{{\bf{w}}_k}} \right|}^2}}}{{\sum\nolimits_{i \in \mathcal{K}/\{k\}} {{{\left| {{\bf{h}}_k^H{{\bf{w}}_i}} \right|}^2}}  + {\sigma_k ^2}}}} \right).
\end{flalign}

Combine \eqref{power} and \eqref{rate}. The EE for the MU-MISO system is expressed as follows:
\begin{flalign}
{\rm EE}\left( {\left\{ {{{\bf{w}}_i}} \right\}} \right) = \frac{{\sum\nolimits_{k \in {\mathcal K}} {{R_k}\left( {\left\{ {{{\bf{w}}_i}} \right\}} \right)} }}{P_{\rm Total}\left( \{{{\bf{w}}_i}\} \right)},
\end{flalign}

Our goal is to maximize the system EE under the constraints of rate requirement of each MU and power budget of the transmitter, which is mathematically formulated as the following optimization problem:
\begin{subequations}\label{p0}
\begin{align}
{\bf P_0:}\mathop {\max }\limits_{\left\{{{{\bf{w}}_i}} \right\}} ~& {\rm{EE}}\left( {\left\{ {{{\bf{w}}_i}} \right\}} \right) \nonumber \\
{\rm s.t.}~&{{R_k}\left( {\left\{ {{{\bf{w}}_i}} \right\}} \right)} \ge \xi_k,\label{cons:A}\\
&\sum\nolimits_{k \in {\mathcal K}}\left\| {{{\bf{w}}_i}} \right\|^2 \le {P_{\max}},\label{cons:B}\\
&{{\bf{w}}_i}\in{\mathbb{C}^{{N_{\rm{T}}}}}, {\forall i,k}\in{\cal K},\label{cons:C}
\end{align}
\end{subequations}
where $\xi_k$ denotes the rate requirement of the $k$-th MU and ${P_{\max}}$ denotes the power budget of the transmitter. 

As a classical beamforming design, the problem $\rm P_0$ as well as its other versions have been widely investigated based on both CVX optimization and DL approaches. However, when it comes to the ultra-dense scenario, the following challenges hinder the development of efficient resource allocation algorithms.  
\begin{itemize}
  \item [\emph{a)}]
 Time-varying CSI: In mobile communication systems, the CSI of MUs changes very fast. Thus, the resource allocation algorithms are require to realize a time response to the input CSI. However, for the CVX-based approach, it relies on time-costly iterative computations as there is no close-formed solution to the problem $\rm P_0$ while for the existing DL approaches, they may fail to yield high-quality solution as they are not designed for ultra-dense scenarios. 
  \item [\emph{b)}]
Scalability to MUs: The memoryless nature makes CVX-based approaches with poor scalability. Existing GNN-based approaches show good scalability for small-scale only, it is challenging to these DL schemes to the ultra-dense scenarios due to greatly increased variables. To facilitate the GNN-base models in ultra-dense scenarios, it is required to either develop powerful mechanisms or bring prior knowledge in the models. However, the former may degrade the inference speed. 
  \item [\emph{c)}]
Various channel conditions and QoS requirements: In ultra-dense scenarios, the MUs are with various channel conditions (reflected by the signal-noise ratio (SNR) or power of AWGN) and QoS requirements (reflected by $\xi_k$), which makes the adaptability be the central consideration of the models. However, those models pay limited attention on the adaptability. 
\end{itemize}

As the DL is task-oriented, it is necessary to develop specified neural networks for the considered problem with large-scale settings. Besides, a trade-off usually exists between the expressive capability and the inference speed of the DL models. Therefore, we intend to leverage the GNN and prior knowledge, the MMSE and HZM schemes, to develop a model-based GNN  to jointly realize real-time and near-optimal inference while being with good scalability to MUs and adaptability to various channel conditions and QoS requirements.




\section{Model-based GNN }

\begin{figure}[t]
\begin{center}
{\includegraphics[ width=.48\textwidth]{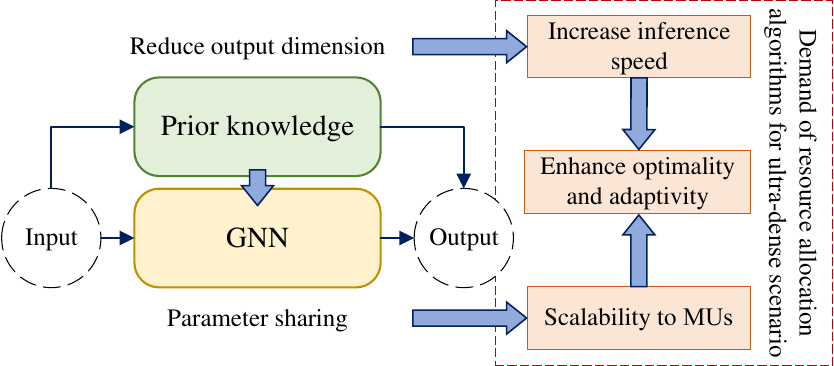}}
\caption{The main idea of the model-based GNN for ultra-dense scenarios.}
\label{mgnn}
\end{center}
\end{figure}

The main idea of the model-based GNN is illustrated in Figure \ref{mgnn}. The prior knowledge is used to ``simplify" the output of the GNN based on its inputs to increase the inference speed. Meanwhile, the parameter sharing in GNN facilitates the scalability to MUs. In addition, by combining prior knowledge and GNN, the feature extracting capability of the DL model is subsequently empowered, which enhances its optimality and adaptivity. In the following, the detailed processes of the model-based GNN including 1) problem reformulation with prior knowledge, 2) architecture of GNN, 3) learning framework are presented.

\subsection{Problem Reformulation with Prior Knowledge}

The main idea of the DL approach is to construct neural networks to learn the (near-)optimal mapping from $\{{\bf h}_k\}$ to $\{ {\bf w}_i\}$. However, the complex-valued vectors of the inputs and outputs complicate the learning task, which may degrade the learning performance and reduce the computational efficiency. Therefore, some prior knowledge can be utilized to reconstruct the mapping with limited performance loss. Particularly, for the $k$-th  beamforming vector, it contains the power part, denoted by $p_k$, and the direction part, denoted by ${{\overline{\bf{w}}}_k}$, i.e.,
\begin{flalign}
{{\bf{w}}_k} = {\sqrt {p_k}}{{\overline{\bf{w}}}_k},~{\left\| {{{\overline {\bf{w}} }_k}} \right\|^2} = 1.
\end{flalign}

As for the direction part, two schemes are adopted, i.e., the classical MMSE scheme and the HZM scheme proposed in \cite{hyb}:

\subsubsection{MMSE scheme} The MMSE scheme fixes the direction part of ${{{\bf{w}}}_k}$ by 
\begin{flalign}\label{MMSE}
{\overline{\bf w}}^{(\rm MMSE)}_k = \frac{{\bf v}_k}{\|{\bf v}_k\|},
\end{flalign}
where ${{\bf v}_k}$ is the $k$-th column of ${\bf V}$ and
\begin{flalign}
    {\bf V} = {\bf G}^H({\bf G}{\bf G}^H + \sigma_k^2{\bf I}_{N}  )^{-1},
\end{flalign}
where ${\bf G}\triangleq[{\bf h}_1^H,{\bf h}_2^H,...,{\bf h}_K^H]$ denotes the conjugate transpose matrix of CSI in the system. 

Then, the problem $\rm P_0$ is reformulated as the following MMSE scheme:
\begin{subequations}
\begin{align}
{\bf P_{1-A}:} \mathop {\max }\limits_{\left\{ p_i  \right\}} ~&  {\rm{EE}}\left( {\left\{ {\sqrt {p_i}}{\overline{\bf w}}^{(\rm MMSE)}_k\right\}} \right) \nonumber \\
{\rm s.t.}~&{{R_k}\left( {\left\{ {\sqrt {p_i}}{\overline{\bf w}}^{(\rm MMSE)}_k \right\}} \right)} \ge \xi_k.\label{cons:p1:A}\\
&\sum\nolimits_{i \in {\mathcal K}} p_i \le {P_{\max}},\label{cons:p1:B}\\
&p_i\ge 0,~i\in {\mathcal K}.
\end{align}
\end{subequations}
It is observed that the output dimension is reduced from complex-valued $KN_{\rm T}$ in the problem ${\rm P}_0$ to real-valued $K$ in the problem ${\rm P}_{\rm 1-A}$.
 
The MMSE scheme is shown to achieve good performance in both the high-SNR and low-SNR region, which is also superior to ZF and MRT schemes. However, the MMSE scheme fixes the directions of beamforming vector.   

\subsubsection{HZM scheme} The HZM scheme set the direction part of ${\bf w}_k$ by a linear combination of the ZF direction and the MRT direction\cite{hyb}, i.e., 
\begin{flalign}\label{hybrid}
{\overline {\bf w}}^{(\rm HZM)}_k \left(\alpha_k\right) = \frac{\alpha_k \frac{{\bf u}_k}{\|{\bf u}_k\|} + \left(1-\alpha_k\right)\frac{{\bf g}_{k}}{\|{\bf g}_{k}\|}}{\|\alpha_k \frac{{\bf u}_k}{\|{\bf u}_k\|} + \left(1-\alpha_k\right)\frac{{\bf g}_{k}}{\|{\bf g}_{k}\|}\|},
\end{flalign}
where $\alpha_k \in [0,1]$ is an optimization variable which denotes the hybrid coefficient, ${\bf u}_k$ or ${\bf g}_k$ is the $k$-th column of ${\bf U}$ or ${\bf G}$, and ${\bf U} = {\bf G}^H({\bf G}{\bf G}^H)^{-1}$.

Instead of fixing the direction part like the MMSE scheme, the HZM scheme allows to fine-tune $\alpha_k$ according to the scenarios. Particularly, in the HZM scheme, $\alpha_k = 1$ represents the ZF scheme while $\alpha_k = 0$ represents the MRT scheme.


Then, the problem $\rm P_0$ is reformulated as the following HZM scheme:
\begin{subequations}
\begin{align}
{\bf P_{1-B}:} \mathop {\max }\limits_{\left\{ p_i,\alpha_i  \right\}} & {\rm{EE}}\left( \left\{{\sqrt {p_i}}{\overline{\bf w}}^{(\rm HZM)}_k\left(\alpha_i\right) \right\}\right) \nonumber \\
{\rm s.t.}~&{{R_k}\left( \left\{ {\sqrt {p_i}}{\overline{\bf w}}^{(\rm HZM)}_k\left(\alpha_i\right) \right\} \right)} \ge \xi_k.\label{cons:p2:A}\\
&\sum\nolimits_{i \in {\mathcal K}} p_i \le {P_{\max}},\label{cons:p2:B}\\
&p_i\ge 0,~\alpha_i\in \left[0,1\right],~i\in {\mathcal K}.\label{cons:p2:c}
\end{align}
\end{subequations}
It is observed that the output dimension is reduced from complex-valued $KN_{\rm T}$ in the problem ${\rm P}_0$ to real-valued $2K$ in the problem ${\rm P}_{\rm 1-B}$. 

\newtheorem{Remark}{Remark}
\begin{Remark}\label{Rem0} 
(Comparison between MMSE and HZM) The output dimension of HZM scheme is two times of that of the MMSE scheme. However, the HZM scheme is not always superior to the MMSE scheme for the task-oriented DL. In some scenarios, the MMSE scheme performs like the optimal design indeed, and the HZM scheme may be inferior to the MMSE scheme due to involving more variables, i.e., $\{\alpha_k\}$, to learn. Nevertheless, the HZM scheme is expected to be applicable to most scenarios due to more degree of freedom in designing beamforming direction. 

\end{Remark}

\subsection{Architecture of GNN}

\begin{table*}[ht]
	\caption{Summary of Main Notations Used in Architecture of GNN}
	\centering
	\begin{tabular}{c|c}
		\hline
		{\bf Notation} & {\bf Definition} \\
		\hline\hline
	${\bf h}_k^{(l)} \in \mathbb{C}^{I^{(l)}}$ &  Input feature of the $k$-th node of the $l$-th CGAL  \\
  \hline
  ${\bf h}_k^{(l+1)} \in \mathbb{C}^{I^{(l+1)} \times D^{(l)}}$ &  Output feature of the $k$-th node of the $l$-th CGAL  \\
  \hline
        $\gamma_{d,i,j}^{(l)} \in {\mathbb R}$ & Attention weight between nodes $i$ and $j$ of $d$-th attention head of the $l$-th CGAL\\
  \hline
        ${\bf a}_d^{(l)}\in {\mathbb C}^{I^{(l+1)}}$ & Learnable vector parameter of $d$-th attention head of the $l$-th CGAL in multi-head attention\\
   \hline
        
        ${\bf W}_{{\rm S},d}^{(l)}, {\bf W}_{{\rm N},d}^{(l)} \in {\mathbb C}^{I^{(l+1)} \times I^{(l)}}$ & Learnable matrix parameter of $d$-th attention head of the $l$-th CGAL in multi-head attention\\
  \hline
        ${\bf m}_{d,i}^{(l)} \in {\mathbb C}^{I^{(l+1)}}$ &
        The received message of $d$-th attention head of the $l$-th CGAL\\
  \hline
        ${\bf W}_{{\rm M},d}^{(l)}\in {\mathbb C}^{I^{(l+1)} \times I^{(l)}}$ & Learnable matrix parameter of $d$-th attention head of the $l$-th CGAL in message aggregation \\
   \hline
        ${\widetilde{\bf h}_k^{(t)} \in \mathbb{C}^{K \times O(t)}}$ & Input feature of the $k$-th node of the $t$-th CFCL \\
   \hline
        ${\Tilde{\bf h}_k^{(t+1)} \in \mathbb{C}^{K \times O(t+1)}}$  &  Output feature of the $k$-th node of the $t$-th CFCL \\
   \hline
        ${\bf W}^{(t)}_{\rm M} \in \mathbb{R}^{ O(t+1) \times O(t)}$  &  Learnable matrix parameter of $t$-th CFCL \\
   \hline
	\end{tabular}
 \label{table:pare}
\end{table*}

The considered MU-MISO system can be represented as an undirected fully-connected graph, as illustrated in Figure \ref{sys}, where the $k$-th node represents the desired link between the transmitter and the $k$-th MU and the corresponding node feature is set by ${\bf h}_k$. Besides, the non-feature edge between nodes represents the existing of the interference relationship.

The GNN is to extract hidden features in the graph-structured input, i.e., $\{{\bf h}_i\}$, to derive (near-)optimal $\{{\bf w}_i\}$, which is denoted by
\begin{flalign}
     \left\{{\bf w}_i\right\} = \Pi_{\bm \theta}\left( \left\{ {\bf h}_i \right\}\right):~\mathbb{C}^{K\times{N_{\rm T}}}
   \to \mathbb{C}^{K\times{N_{\rm T}}},
\end{flalign}
where $\Pi_{\bm \theta}$ represents the mapping with  ${\bm \theta}$ being the learnable parameters. 

\begin{figure*}[t]
\begin{center}
{\includegraphics[ width=0.9\linewidth]{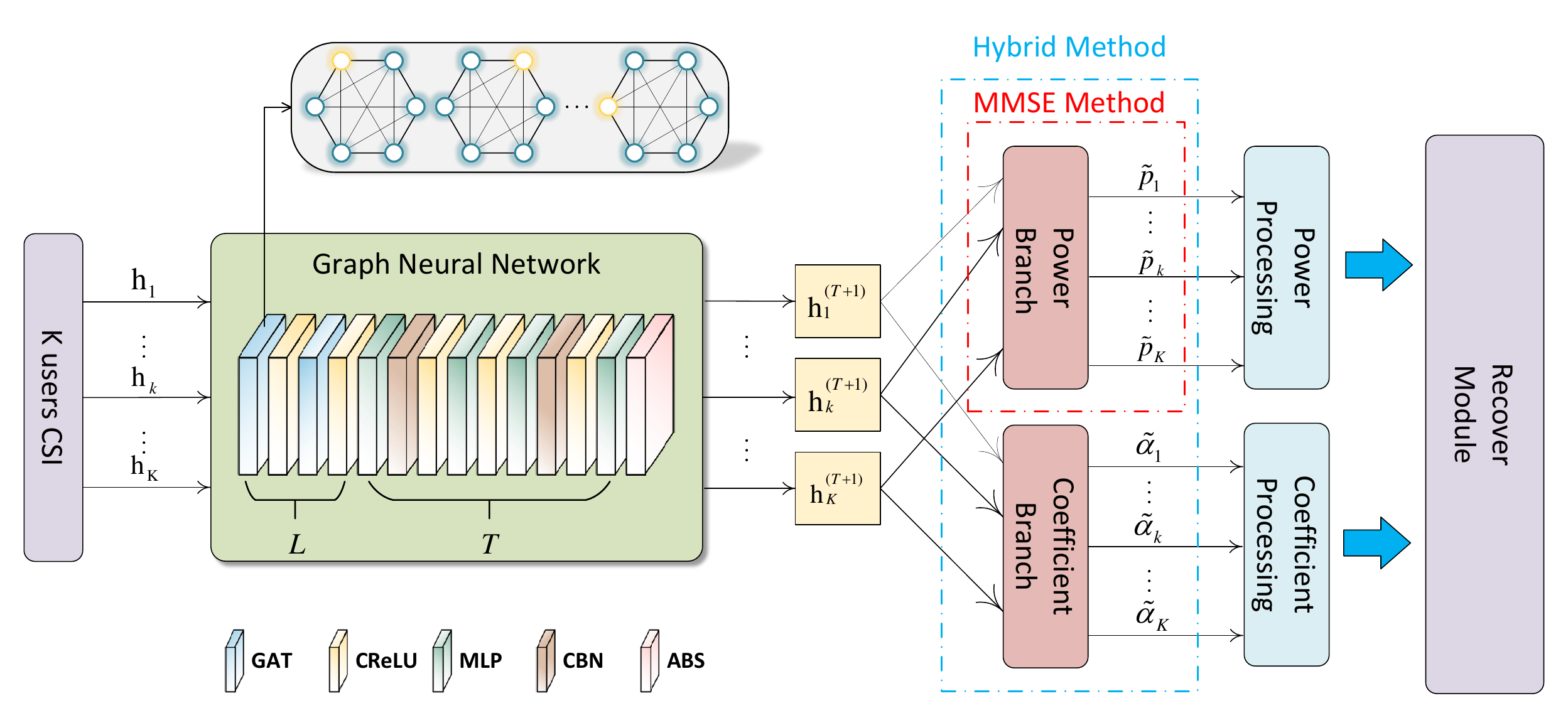}}
\caption{The illustration of the architecture of the model-based GNN.}
\label{model}
\end{center}
\end{figure*}

The desired mapping is realized by the proposed GNN which includes three components, i.e., $L$ complex graph attention layers (CGALs) for feature extracting, one MLP for yielding the require output according to the prior knowledge and one post-processing module to select the scheme and recover the output to the desired beamforming vector. The proposed GNN is illustrated in Figure \ref{model}. Note that the MMSE and HZM schemes share the same CGALs and parts of MLP. The detailed processes are given as follows with the main notations summarized in Table \ref{table:pare}.

\subsubsection{CGAL} The CGALs utilize the message passing to extract the inter-link relationship (e.g., interference) to update the node features. It is important to distinguish the impact levels of other nodes for a given node. Therefore, the multi-head attention is adopted to compute the weights of messages. The key steps, including attention, aggregation and update, of the $l$-th CGAL are given as follows. In brief, the CGAL is comprised of three steps, i.e., multi-head attention, message aggregation and feature update.  

\textbf{Multi-head attention}: Denote ${\bf h}_k^{(l)} \in \mathbb{C}^{I^{(l)}}$ and ${\bf h}_k^{(l+1)} \in \mathbb{C}^{I^{(l+1)} \times D^{(l)}}$ by the input and output features\footnote{For the $i$-th node, $L$ CGALs map ${\bf h}_i^{(1)}={\bf h}_i \in \mathbb{C}^{K\times{N_{\rm T}}}$ to ${\bf h}_i^{(L+1)}\in \mathbb{C}^{I^{(L+1)} \times D^{(L)}}$. } of the $k$-th node of the $l$-th CGAL, respectively, where $I^{(l)}$ and ${I^{(l+1)}}$ denote the corresponding dimensions and $D^{(l)}$ denotes the number of attention heads. 

For the $i$-th node, it assigns an attention weight denoted by ${\gamma_{d,i,j}^{(l)}}$ to its $j$-th neighboring node via the $d$-th attention head, which is given by 
\begin{flalign}\label{att}
    &\gamma_{d,i,j}^{(l)} = \\
    &\frac
    {\exp  \left( \left|  {\bf a}_d^{(l)^T} {\rm {\mathbb C}LeakyReLU}\left( {\bf W}_{{\rm S},d}^{(l)}{\bf h}_{i}^{(l)} + {\bf W}_{{\rm N},d}^{(l)}{\bf h}_{j}^{(l)} \right) \right|  \right)}
    {\sum\limits_{k \in {\cal K}} \exp  \left( \left|  {\bf a}_d^{(l)^T} {\rm {\mathbb C}LeakyReLU}\left( {\bf W}_{{\rm S},d}^{(l)}{\bf h}_{i}^{(l)} + {\bf W}_{{\rm N},d}^{(l)}{\bf h}_{k}^{(l)} \right) \right|  \right)},\nonumber
\end{flalign}
where  ${\bf a}_d^{(l)}\in {\mathbb C}^{I^{(l+1)}}$, ${\bf W}_{{\rm S},d}^{(l)}\in {\mathbb C}^{I^{(l+1)} \times I^{(l)}}$ and ${\bf W}_{{\rm N},d}^{(l)}\in {\mathbb C}^{I^{(l+1)} \times I^{(l)}}$ denote the learnable parameters of the $d$-th attention head in the $l$-th CGAL, and ${\rm {\mathbb C}LeakyReLU}(\cdot)$ denotes the complex-valued leaky rectified linear unit activation function.  Note that in \eqref{att}, the self-loop is included to enhance the learning stability. 


\textbf{Message aggregation}: Following the $d$-th attention head, the $i$-th node takes attention-assisted aggregation over its neighbouring nodes as well as itself. The received weighted message is given by
\begin{flalign}
    {\bf m}_{d,i}^{(l)}=\mathop {\rm Sum}\limits_j \left(\left\{ \gamma_{d,i,j}^{(l)}{\bf W}_{{\rm M},d}^{(l)}{\bf h}_{j}^{(l)} \right\}_j\right)\in{\mathbb{C}^{I^{(l+1)}}},~ j \in {\cal K},
\end{flalign} 
where ${\bf W}_{{\rm M},d}^{(l)}\in {\mathbb C}^{I^{(l+1)} \times I^{(l)}}$ denotes the learnable parameters of the aggregation, and ${\rm Sum}(\cdot)$ denotes the sum operation.

\textbf{Feature update}: With $D^{(l)}$ aggregated messages, the $i$-th node updates its node feature as the node-level output of $l$-th CGAL, which is given by
\begin{flalign}
    {\bf h}_i^{(l+1)}={\rm Com}\left(\left\{{\bf m}_{d,i}^{(l)}\right\}_d\right),
\end{flalign}
where ${\rm Com}(\cdot)$ denotes the concatenation operation.


\subsubsection{MLP}

The MLP decodes the extracted complex node features, i.e., $\{ {\bf h}_i^{(L+1)} \}$ as real-valued ${\left\{p_i \in  \mathbb{R} \right\}}$ for the MMSE scheme and real-valued ${\left\{p_i \in  \mathbb{R},\alpha_i  \in  \mathbb{R} \right\}}$ for the HZM scheme. The MLP is realized by $T$ complex fully-connected layers (CFCLs). 

Denote that ${\{\widetilde{\bf h}_k^{(t)}\} \in \mathbb{C}^{K \times O(t)}}$ and ${\{\Tilde{\bf h}_k^{(t+1)}\} \in \mathbb{C}^{K \times O(t+1)}}$ as the input and output of the $t$-th ($t\in\{1,...,T\}$) CFCL. The computation of the $t$-th CFCL is given by
\begin{flalign}
    \widetilde{\bf h}_k^{(t+1)}=& {\mathbb C}{\rm ReLU}
    \left({\bf W}^{(t)}_{\rm M}\left(\operatorname{Re} \left(\widetilde{\bf h}_k^{(t)} \right)-\operatorname{Im}\left( \widetilde{\bf h}_k^{(t)} \right)\right)\right. +\\
    &\left. j\left({\bf W}^{(t)}_{\rm M}\left(\operatorname{Im}\left( \widetilde{\bf h}_k^{(t)} \right)+\operatorname{Re}\left( \widetilde{\bf h}_k^{(t)} \right)\right)\right)\right)\nonumber,
\end{flalign}
where {${\bf W}^{(t)}_{\rm M} \in \mathbb{R}^{ O(t+1) \times O(t)}$} represents the learnable parameters of the $t$-th CFCL. 

Besides, an additional process is applied to the output of the $T$-th CFCL, i.e.,
\begin{flalign}
[{\widetilde p}_k ]^T ~{\rm or}~ [\widetilde{p}_k, \widetilde{\alpha}_k]^T= {\rm Abs}\left({\rm Re}\left(\widetilde{\bf h}_k^{(T+1)}\right)\right).
\end{flalign}

For both the MMSE and HZM schemes, the output $\{{\widetilde{p}_k}\}$ are adjusted  to satisfy the constraint of power budget \eqref{cons:p1:B} by the following activation function, i.e.,
\begin{flalign}\label{af}
{p_k} = \left\{ \begin{array}{l}
\widetilde{p}_k,~\sum\nolimits_{i=1}^K \widetilde{p}_k \le {P_{\max}},\\
\frac{{\widetilde{p}_k}}{{\sum\nolimits_{i = 1}^K {\widetilde{p}_k} }}{P_{\max }},~\sum\nolimits_{i=1}^K \widetilde{p}_k> {P_{\max}}.
\end{array} \right.
\end{flalign}

In addition, for the HZM scheme, the output $\{\widetilde\alpha_k\}$ are adjusted to satisfy  \eqref{cons:p2:c} by the Sigmoid function, i.e., 
\begin{flalign}
\alpha_k = \frac{1}{1+\exp\left({\widetilde\alpha_k}\right)}.
\end{flalign}

\subsubsection{Post-processing Module}

\begin{figure*}
\begin{flalign}\label{x}
{\bf z} =\mathop{\arg\max}\limits_{{\bf z}=[{p}_1,...,{p}_K]^T ~{\rm or}~ [{p}_1,...,{p}_K,\alpha_1,...,\alpha_K]^T}\left\{{\partial ^{\left( {{\rm{MMSE}}} \right)}}{\rm{EE}}\left( {\left\{ {\sqrt {{ p}_i}}{\overline{\bf w}}^{(\rm MMSE)}_k\right\}} \right) ,{\partial ^{\left( {{\rm{HZM}}} \right)}}{\rm{EE}}\left( \left\{{\sqrt {{ p}_i}}{\overline{\bf w}}^{(\rm HZM)}_k\left({\alpha}_i\right) \right\}\right)  \right\}
\end{flalign}
\hrule
\end{figure*}

The post-processing module includes two processes, i.e., the scheme selection module and the beamforming recover module. 

The scheme selection module is to select one scheme from the two schemes, namely the MMSE and HZM schemes, with the better performance according to the given input (which reflects the channel conditions). As shown in Fig \ref{model}, the MMSE scheme and the HZM scheme share the CGALs but adopt different MLPs. Therefore, the scheme selection module can be realized via \eqref{x}, where $\bf z$ denotes the output of the scheme selection module, which can be $[{p}_1,...,{p}_K]^T\in {\mathbb R}^K$ for the MMSE scheme  or $[{p}_1,...,{p}_K,\alpha_1,...,\alpha_K]^T\in {\mathbb R}^{2K}$ for the HZM scheme. Here, ${\partial ^{( {{\rm{MMSE}}})}}$/${\partial ^{( {{\rm{HZM}}})}}=1$ if and only if $\{p_i\}$/$\{p_i,\alpha_i\}$ is feasible to the problem ${\rm P_{1-A}/P_{1-B}}$; otherwise ${\partial ^{( {{\rm{MMSE}}})}}$/${\partial ^{( {{\rm{HZM}}})}}=0$. 

The beamforming recover module is to yield the desired beamforming vectors based on the selected scheme, i.e., 
\begin{align}
    &{{\bf w}}_k  = \left\{ \begin{array}{l} 
    \sqrt{p_k} {\overline{\bf w}}^{(\rm MMSE)}_k ,~ {if}~{{\bf z}\in {\mathbb R}^K}\\
    \sqrt{p_k} {\overline {\bf w}}^{(\rm HZM)}_k \left(\alpha_k\right),~ {{if}~{{\bf z}\in{\mathbb R}^{2K}}}
    \end{array} \right.,~\forall k.\nonumber
\end{align}

As a result, the mapping from $\{{\bf h}_i\}$ to $\{{\bf w}_i\}$ is constructed via the GNN with the learning parameters ${\bm \theta} = \{{\bf a}_d^{(l)}, {\bf W}_{{\rm S},d}^{(l)}, {\bf W}_{{\rm N},d}^{(l)}, {\bf W}_{{\rm M},d}^{(l)}, {\bf W}^{(t)}_{\rm M}\}$. Note that all nodes share ${\bm \theta}$ during the node feature updating, and the permutation invariant and equivalence of GNN are guaranteed. 


\subsection{{Learning Framework}}

The goal of the model-based GNN is to obtain the (near-)optimal solution to the problem $\rm P_0$. Therefore, an efficient learning framework is required to train the GNN. 

The unsupervised learning is adopted to alleviate the high overhead to obtain the labelled training set, especially for ultra-dense scenarios. Note that the unsupervised learning does not explicitly yield the training efficiency in general. However, when it comes to the optimization problem, the convergence of the (constrained) objective function can be adopted as the indicator of training efficiency. Therefore, the loss function is designed as 
\begin{flalign}
{\cal L}_N\left( {\bm\theta} \right) = &\frac{1}{N} \sum\nolimits_{n=1}^N \left(-{\rm EE}\left( {\left\{{\bf w}_k \right\}}^{(n)} \right) + \right. \label{loss_f}\\
&\left. \sum\nolimits_{i=1}^K    \lambda {\rm ReLU}\left( \xi_k - {{R_k}\left( {\left\{{\bf w}_k \right\}} \right)^{(n)}} \right)\right)\nonumber,
\end{flalign}
where $N$ denotes the number of samples (indexed by $n$) in a batch, and  $\lambda>0$ is a hyperparameter which denotes the penalty factor. With the penalty term, the loss function shall be increased if an infeasbile solution is used \eqref{cons:A}, which can conflict with the minimization of the loss function.

Combine the parameter sharing of the GNN and the unsupervised learning, we adopt two training strategies, i.e., the constant-input training strategy and the various-input training strategy:
\begin{itemize}
  \item [\emph{a)}]
For the constant-input training strategy, the numbers of MUs of the training samples are identical. 
  \item [\emph{b)}]
For the various-input training strategy, the numbers of MUs of the training samples can be different. 
\end{itemize}
The motivation for the latter one is to enhance the stability of the GNN in various application scenarios. The reason is that in ultra-dense scenarios, the model-based GNN may encounter plenty of unseen problem sizes. Although the GNN is adaptable to  different input dimensions, the constraints may not be satisfied at a tolerable level for some unseen problem sizes. Especially, in the considered problem, the QoS requirement is considered for each MU. A poor generalization performance shall induce infeasible QoS requirement which greatly degrades the user experience. With the various-input training strategy, the model-based GNN can learn more types of data structures during the training phase. Nevertheless, one drawback is that with a fixed number of training samples, the training samples for a specific scenario are decreased in the various-input training strategy than the constant-input training strategy, which may induce some performance loss for the scenario.

\section{Simulation Results}

This section provides numerical results to validate the performace of proposed model-based GNN for large-scale wireless optimization. Particularly, the scheme selection module in the post-processing module is inactivated in order to compare the MMSE scheme and the HZM scheme unless specified. The structure of the model-based GNN under test is given in Table \ref{struct}. 

\begin{table}[t]
    \centering
    \caption{The Structure of GNN}
    \begin{tabular}{@{}c|c|c|c|c|c|c@{}}
    \hline
        No. & Layer & Input & Output & AMs & CReLU & CBN \\
    \hline
     \hline
        1 & CGAL & $N_T$ & 64 & 20 & \checkmark & $\times$ \\
    \hline
        2 & CGAL & 1,280 & 512 & 20 & \checkmark & $\times$ \\
    \hline
        3 & CFCL & 10,240 & 512 & - & \checkmark & \checkmark \\
    \hline
        4 & CFCL & 512 & 128 & - & \checkmark & \checkmark \\
    \hline
        5 & CFCL & 128 & $M_n$ & - & $\times$ & $\times$ \\
    \hline
    \end{tabular}\label{struct}
    \begin{tablenotes}
    \footnotesize
        \item{AMs: Number of attention mechanisms} \\
        \item{$M_n$: $M_n=1$ for MMSE scheme  and $M_n=2$ for HZM scheme.}
    \end{tablenotes}
\end{table}

\subsection{Simulation setting}

\begin{table}[t]
\centering
\caption{Simulation Parameters Used in Simulation}
\begin{tabular}{c|c }
\hline
{\bf Notation}               & {\bf Values}               \\ \hline
\hline
Number of MUs & ${K \in \{15, 20, 30, 35, 40, 45, 50\}}$ \\ \hline
Number of antennas               & ${ N_{\rm T} = 64 }$                \\ \hline
Average $\raise0.5ex\hbox{{1}} \!\mathord{\left/
 {\vphantom {{} {}}}\right.\kern-\nulldelimiterspace}
\!\lower0.5ex\hbox{{\rm SNR}}$ & $\Gamma\in\{0.1, 0.25, 0.5, 0.75, 1\}$ \\ \hline
Rate requirement & ${\xi \in \{ 0.5, 0.75, 1 \}}$ \\ \hline
Power budget   & ${P_{\rm max}} = 1$ W   \\ \hline 
Circuit power  &  ${P_{\rm C} = 0.5}$ W \\ \hline
\end{tabular}
\label{Simulation setting}
\end{table}

\begin{table}[ht]
\centering
\caption{Datasets.}
\begin{tabular}{c|c|c|c|c|c|c}
\hline
No. &$N_{\rm T}$ &$K$ & $\Gamma$ & ${\xi}$& Size& Type\\
 \hline
 \hline
 1   &64&15&0.5&1& 2,000& B\\
 \hline
2   &64&20&0.5&1&  2,000& B\\
 \hline
 3   &64&30&0.1&0.5& 12,000& A\\
  \hline
4   &64&30&0.1&0.75& 12,000& A\\
  \hline
5   &64&30&0.1&1& 12,000& A\\
  \hline
6   &64&30&0.25&0.5& 12,000& A\\
  \hline
7   &64&30&0.25&0.75& 12,000& A\\
  \hline
8   &64&30&0.25&1& 12,000& A\\
  \hline
9   &64&30&0.5&0.5& 12,000& A\\
  \hline
10   &64&30&0.5&0.75& 12,000& A\\
  \hline
11   &64&30&0.5&1& 12,000& A\\
  \hline
12   &64&30&0.75&0.5& 12,000& A\\
  \hline
13   &64&30&0.75&0.75& 12,000& A\\
  \hline
14   &64&30&0.75&1& 12,000& A\\
  \hline
15   &64&30&1&0.5& 12,000& A\\
  \hline
16   &64&30&1&0.75& 12,000& A\\
  \hline
17   &64&30&1&1& 12,000& A\\
  \hline
18   &64&35&0.5&1& 2,000& B\\
 \hline
19   &64&40&0.5&1& 2,000& B\\
 \hline
20   &64&45&0.5&1& 2,000& B\\
 \hline
21   &64&50&0.5&1& 2,000& B\\
 \hline
22   &64&$\{30,40,50\}$&0.5&1& 10,000& C\\
 \hline
\end{tabular}
 \begin{tablenotes}
        \footnotesize
        \item {Type A: The training set and test set are all included.}
        \item {Type B: Only test set is included.}
        \item {Type C: Only training set is included.}
\end{tablenotes}
\label{dataset}
\end{table}

\subsubsection{Simulation scenario} The number of antennas is set as $N_{\rm T}=64$. The number of MUs is set as $K\in\{15,20,30,35,40,45,50\}$. The power budget and the constant circuit power are set as $P_{\rm C}=0.5$ W and $P_{\rm max}=1$ W. The rate requirements of $K$ MUs are identically set as $\xi_1=...=\xi_K=\xi \in \{0.5,0.75,1\}$ bit/s/Hz. The CSI of each MU includes both large-scale path loss and the small-scale fading. The path-loss model is based on $10^{-(140.7+36.7\log_{10}({\widetilde d}))/10}$, where ${\widetilde d}$ (in kilometer) denotes the distance between the BS and the MU. Rayleigh fading is adopted for the small-scale fading. The noise power spectrum is $162$ dBm/Hz and the bandwidth is $10$ MHz. We consider several channel conditions with average $\raise0.5ex\hbox{{1}} \!\mathord{\left/
 {\vphantom {{} {}}}\right.\kern-\nulldelimiterspace}
\!\lower0.5ex\hbox{{\rm SNR}}$ denoted by $\Gamma$ being $\{0.1, 0.25, 0.5, 0.75, 1\}$. The simulation parameters are summarized in Table \ref{Simulation setting}.

\subsubsection{Dataset} 
Each training sample includes $K$ channel vectors, i.e.,  $\{ {\bf h}_{i} \}$ while each test sample includes $K$ channel vectors and a label which represents the corresponding maximum EE.  In this work, we build $22$ datasets with different parameter settings as shown in Table\ref{dataset}. Particularly, the datasets are categorized into three types. 
\begin{itemize}
    \item{Type A:} This type of dataset includes $10,000$ training  samples and $2,000$ test samples, and the training sample and the test sample have the identical parameter setting.
    \item{Type B:} This type of dataset includes $2,000$ test samples only, which is used to test the model-based GNN with the unseen problem sizes. 
    \item{Type C:} This type of dataset includes $10,000$ training samples only, and the values of $K$ of training samples can be  different.
\end{itemize}

\subsubsection{Baseline}
In order to evaluate the model-based GNN numerically, the following three baselines are also considered, i.e.,
\begin{itemize}
\item \textbf{CVX-based approach:} A traditional CVX optimization algorithm is adopted, e.g., a single-loop solution approach in Appendix \ref{slsa}.
\item \textbf{Model-based  MLP:} A basic feed-forward  neural network based on CFCLs is adopted \cite{MLP1} to replace the CGALs in the model-based GNN. 
\item \textbf{Model-based  CNN:} A basic CNN is adopted to replace the CGALs in the model-based GNN. The input of CNN consists of three channels, namely the absolute values, the real parts and the imaginary parts of ${\bf H}\triangleq[{\bf h}_1,...,{\bf h}_K]^T\in {\mathbb C}^{K\times {N_{\rm T}}}$ \cite{CNN1}. For each channel, a 2D (complex) convolutional kernel with size of $K\times {N_{\rm T}}$ is adopted. 
\end{itemize}

\subsubsection{Implementation details} As mentioned, all neural networks are trained in an unsupervised learning manner. The Adam\cite{adam} is adopted as the optimizer. The initialized choice of the learning rate ${10^{-3}}$. The batch size is set as $25$ and the maximum epoch of training is set as $400$. The learnable weights with the best performance are used as the training result. The CVX formulations and the NNs are respectively processed by CVX solver {\tt SeDuMi} under Mathworks MATLAB R2021b and Python 3.8.18 with Pytorch 1.10.0 on a computer with Intel(R) Core(TM) i9-12900K CPU and one NVIDIA RTX 3090 GPU (24 GB of memory).

\subsubsection{Performance metrics}  The following metrics are considered to evaluate the performance of the model-based GNN from the perspectives of model effectiveness and task requirement. 
\begin{itemize}
  \item [\emph{a)}] 
  \emph{Optimality performance:} The average ratio of the achievable EE (with feasible solutions) by the neural network to the maximum EE on the test set with identical settings with the training set.
    \item [\emph{b)}]
 \emph{Scalability performance:} The average ratio of the achievable EE (with feasible solutions) by the neural network to the maximum EE on the test set with different settings from the training set.
  \item [\emph{c)}] 
  \emph{Feasibility rate:} The percentage of the feasible solutions to the considered problem by the neural network.
  \item [\emph{d)}]
 \emph{Inference time:} The average running time required to calculate the feasible solution with the given CSI by the neural network or traditional CVX-based approach.  
\end{itemize}
Note that we evaluate the optimality performance on the test set of the type-A dataset while the scalability performance on the type-B dataset. Besides, the feasibility rate is only related to the constraint \eqref{cons:A} which stands for the QoS performance, since the constraint \eqref{cons:B} is guarantee to be satisfied by \eqref{af}.

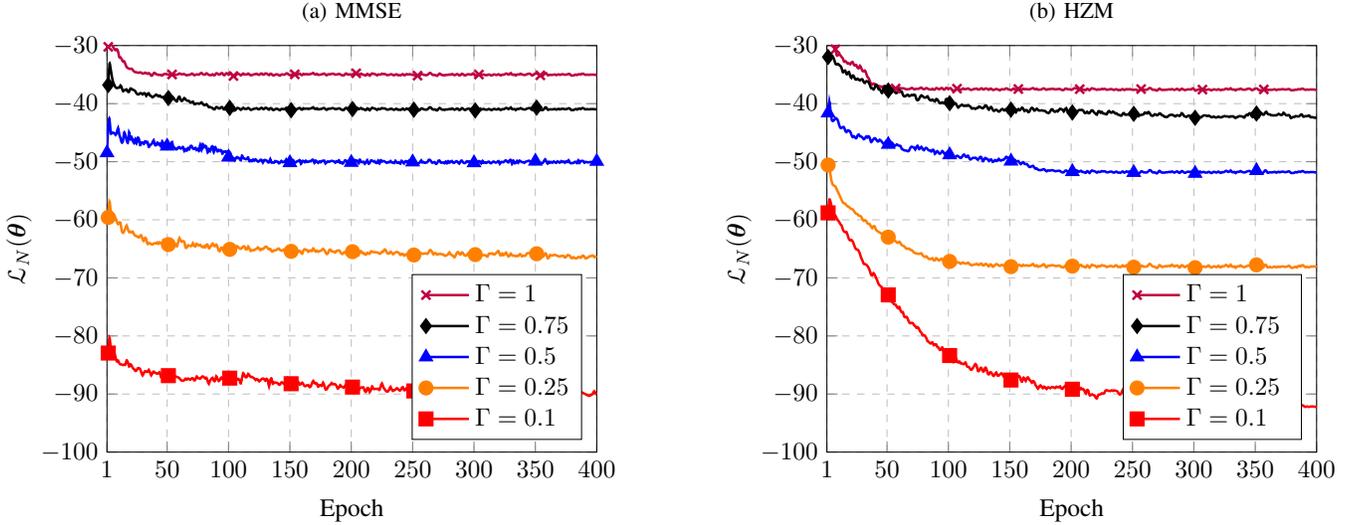
\begin{figure*}[t]
    \centering
\subfigure{
\begin{tikzpicture}[scale = 0.95]
\label{strategy:a}
\begin{axis}[
    title={(a) MMSE},
    title style={font=\small},
    xlabel={Epoch},
    ylabel={${\cal L}_N( {\bm\theta} )$},
    xmin=1, xmax=400,
    ymin=-100, ymax=-30,
    xtick={1,50,100,...,400},
    ytick={-30,-40,...,-100},
    legend pos=south east,
    legend style={
    nodes={right,align=left}},
    xmajorgrids=true,
    ymajorgrids=true,
    grid style=dashed,
]

\addplot[
    color=purple,
    smooth,
    line width = 1.0pt, 
    mark=x,
    mark size=2.5pt,
    mark indices={1,50,100,...,400}
    ]
    table[
        col sep=comma,
        x = Epoch,
        y = Smoothed_Train_Loss
    ]{data/MMSE_1.csv};
\addplot[
    color=black,
    smooth,
    line width = 1.0pt, 
    mark size=2.5pt,
    mark=diamond*, 
    mark indices={1,50,100,...,400}
    ]
    table[
        col sep=comma,
        x = Epoch,
        y = Smoothed_Train_Loss
    ]{data/MMSE_075.csv};
\addplot[
    color=blue,
    smooth,
    line width = 1.0pt, 
    mark=triangle*, 
    mark size=2.5pt,
    mark indices={1,50,100,...,400}
    ]
    table[
        col sep=comma,
        x = Epoch,
        y = Train_Loss
    ]{data/MMSE_05.csv};
\addplot[
    color=orange,
    smooth,
    line width = 1.0pt, 
    mark=*, 
    mark size=2.5pt,
    mark indices={1,50,100,...,400}
    ]
    table[
        col sep=comma,
        x = Epoch,
        y = Smoothed_Train_Loss
    ]{data/MMSE_025.csv};
\addplot[
    color=red,
    smooth,
    line width = 1.0pt, 
    mark=square*, 
    mark size=2.5pt,
    mark indices={1,50,100,...,400}
    ]
    table[
        col sep=comma,
        x = Epoch,
        y = Smoothed_Train_Loss
    ]{data/MMSE_01.csv};
    \legend{$\Gamma = 1$,$\Gamma = 0.75$,$\Gamma = 0.5$,$\Gamma = 0.25$,$\Gamma = 0.1$}
\end{axis}
\end{tikzpicture}
}
\hfill
\subfigure{
\begin{tikzpicture}[scale = 0.95]
\label{strategy:b}
\begin{axis}[
    title={(b) HZM},
    title style={font=\small},
    xlabel={Epoch},
    ylabel={${\cal L}_N( {\bm\theta} )$},
    xmin=1, xmax=400,
    ymin=-100, ymax=-30,
    xtick={1,50,100,...,400},
    ytick={-30,-40,...,-100},
    legend pos=south east,
    legend style={
    nodes={right, align=left}},
    xmajorgrids=true,
    ymajorgrids=true,
    grid style=dashed,
]

\addplot[
    color=purple,
    smooth,
    line width = 1.0pt, 
    mark=x,
    mark size=2.5pt,
    mark indices={1,50,100,...,400}
    ]
    table[
        col sep=comma,
        x = Epoch,
        y = Smoothed_Train_Loss
    ]{data/Hybrid_1.csv};
    
\addplot[
    color=black,
    smooth,
    line width = 1.0pt, 
    mark=diamond*, 
    mark size=2.5pt,
    mark indices={1,50,100,...,400}
    ]
    table[
        col sep=comma,
        x = Epoch,
        y = Smoothed_Train_Loss
    ]{data/Hybrid_075.csv};
    
\addplot[
    color=blue,
    smooth,
    line width = 1.0pt, 
    mark=triangle*, 
    mark size=2.5pt,
    mark indices={1,50,100,...,400}
    ]
    table[
        col sep=comma,
        x = Epoch,
        y = Smoothed_Train_Loss
    ]{data/Hybrid_05.csv};
\addplot[
    color=orange,
    smooth,
    line width = 1.0pt, 
    mark=*, 
    mark size=2.5pt,,,
    mark indices={1,50,100,...,400}
    ]
    table[
        col sep=comma,
        x = Epoch,
        y = Smoothed_Train_Loss
    ]{data/Hybrid_025.csv};
\addplot[
    color=red,
    smooth,
    line width = 1.0pt, 
    mark=square*, 
    mark size=2.5pt,,,
    mark indices={1,50,100,...,400}
    ]
    table[
        col sep=comma,
        x = Epoch,
        y = Smoothed_Train_Loss
    ]{data/Hybrid_01.csv};

    \legend{$\Gamma = 1$, $\Gamma = 0.75$,$\Gamma = 0.5$,$\Gamma = 0.25$ , $\Gamma = 0.1$}

\end{axis}
\end{tikzpicture}
}

    \caption{Convergence behavior of the model-based GNN with the MMSE and HZM scheme. }
    \label{train}
\end{figure*}

\subsection{Convergence behavior}

Figures \ref{train}(a) and \ref{train}(b) show the convergence behavior of the model-based GNN with the MMSE and HZM schemes during the training phase, respectively. The results of $\Gamma \in \{0.1, 0.25, 0.5, 0.75, 1\}$ are shown. It is observed that the model-based GNN converges for the considered cases while the model-based GNN with the MMSE scheme converges faster. Besides, $\Gamma$ is shown to have a notable impact on the convergence speed and stability of the model-based GNN with the HZM scheme. Particularly, the  model-based GNN converges faster in the low-SNR region than that in high-SNR region. The reason may be that involving the  hybrid coefficients complicates the training process.

\subsection{Comparison between MMSE and HZM}
In this subsection, we evaluate the model-based GNN via constant-input training strategy in terms of the optimality performance and inference time. 

Table \ref{base experiment} shows that optimality performance of the model-based GNN with the MMSE and HZM schemes  under different $\Gamma$ and $\xi$. The feasibility rate for all simulated cases is $100\%$. It is observed that both MMSE and HZM schemes achieve reasonably good EE performance for different $\Gamma$ and $\xi$. Besides, $\Gamma$ has a great impact on the scheme selection. Particularly, the MMSE scheme performs better in the low-SNR and high-SNR regions while the HZM scheme is more preferable for other cases, especially for ${\Gamma = 0.5}$, which validates the effectiveness of learning the hybrid coefficient in \eqref{hybrid}. The result is consistent with the analysis in Remark \ref{Rem0}.


\begin{table}[ht]
\centering
\caption{Optimality performance of the model-based GNN with $K_{\rm Tr} = K_{\rm Te}=30$.}
\begin{tabular}{c|c|l|l|l}
\hline
\multicolumn{1}{l|}{Method} & \diagbox{$\Gamma$}{${\xi}$} & \multicolumn{1}{c|}{0.5} & \multicolumn{1}{c|}{0.75} & \multicolumn{1}{c}{1} \\ \hline
\multirow{5}{*}{MMSE}       
& 0.1     & \textbf{81.68\%} & \textbf{81.68\%} & \textbf{81.63\%}   
\\ \cline{2-5} 
& 0.25    & 89.72\% & 89.73\% & 87.06\%    
\\ \cline{2-5} 
& 0.5     & 94.05\% & 93.65\% & 94.71\%              
\\ \cline{2-5} 
& 0.75    & 96.66\%  & 96.81\% & 95.06\%              
\\ \cline{2-5} 
& 1     & \textbf{98.39\%}  & \textbf{98.83\%}   & \textbf{97.62\%}              
\\ \hline
\multirow{5}{*}{HZM}
& 0.1     & 80.43\% & 80.27\% & 80.02\%   
\\ \cline{2-5} 
& 0.25    & \textbf{89.88\%} & \textbf{90.71\%} & \textbf{90.91\%}    
\\ \cline{2-5} 
& 0.5     & \textbf{98.49\%} & \textbf{98.32\%} & \textbf{98.32\%}              
\\ \cline{2-5} 
& 0.75    & \textbf{97.87\%}  & \textbf{97.10\%} & \textbf{98.79\%}              
\\ \cline{2-5} 
& 1     & 94.72\%  & 94.13\%   & 94.06\%              
\\ \hline
\end{tabular}\label{base experiment}
     \begin{tablenotes}
        \footnotesize
         \item{$K_{\rm Tr}$: Value of $K$ in the training set.}
        \item{$K_{\rm Te}$: Value of $K$ in the test set.}
        \end{tablenotes}
\end{table}

Table \ref{time0} shows the average inference time of CVX-based approach, model-based MLP, model-based  CNN and model-based  GNN over $2,000$ test samples with $N_{\rm T}=64$ and $K=30$ (Dataset No. 11) . It is observed that the inference time can be greatly reduced by the neural networks compared with the CVX-based approach. Besides, the MLP is with the fastest inference speed due to its simplest architecture. The GNN is shown to be more computationally efficient than the CNN, as the input of GNN is CSI while the input of CNN consists of three channels. Besides, the parameter sharing in GNN reduces the number of learnable parameters which are independent of $K$  while the convolutional kernel with size of $K\times N_{\rm T}$ in CNN requires huge parameters which increases with $K$. Moreover, involving the hybrid coefficients $\{\alpha_k\}$ has limited impact on the inference time, and the millisecond-level inference for both MMSE and HZM schemes facilitates the scheme selection. Combining Table \ref{base experiment} and Table  \ref{time0}, the model-based GNN is shown to be able to make a near-optimal and real-time response to the time-varying CSI.

\begin{table*}[ht]
  \centering
    \caption{Average inference time over the test set of Dataset No. 11.}
    \begin{tabular}{c || c|c|c|c|c|c|c}
      \hline
      \multirow{2}*{Approach} & \multirow{2}*{CVX} & \multicolumn{2}{c|}{MLP} & \multicolumn{2}{c|}{CNN} & \multicolumn{2}{c}{GNN}\\
      \cline{3-8}
       &  & MMSE & HZM & MMSE & HZM & MMSE & HZM \\
      \hline
      Inference time & 11.4 s & 1.60 ms & 1.59 ms & 25.4 ms & 25.4 ms & 4.94 ms & 4.99 ms\\
      \hline
    \end{tabular}
    \label{time0}
\end{table*}



\subsection{Scalability of Model-based GNN}
In this subsection, we evaluate the scalability performance of model-based GNN via the constant-input training strategy, since in ultra-dense scenarios, the number of MUs dynamically changes with time and the channel conditions and QoS requirement varies.

Table\ref{user scale} shows the optimality performance, scalability performance to the number of MUs and feasibility rate of the MLP, CNN and GNN. It is observed that all neural networks achieves good optimality performance with $100\%$ feasibility rate, while GNN with HZM outperforms other baselines. Besides, the feature extractor in CNN and GNN helps them to achieve better optimality performance than MLP. For the scalability performance, the neural networks trained with $K=30$ (Dataset No. 11) are test with $K\in\{15,20,30,35,40,45,50\}$ (Dataset No. 1, 2, 9, 18, 19, 20, 21). Note that due to the fixed input dimension, the MLP and CNN only takes input of CSI with $K\le 30$ by masking some input port by zero. Although the MMSE scheme achieves better performance than the HZM scheme with $K\in\{40,45\}$, the HZM scheme is with a high feasibility rate. Besides, the MMSE scheme fails to yield feasible solution with $K=50$ while the HZM scheme still works. Therefore, the HZM scheme is with better stability.

\begin{table*}[t]
    \centering
    \caption{Adaptability to channel conditions and QoS requirement: $K_{\rm Te}=30$ and $ K_{\rm Te}\in\{15,20,30,35,40,45,50\}$}
    \begin{tabular}{c|c|cc|cc|cc}
\hline
    \multirow{2}{*}{Method} & \multirow{2}{*}{$K_{\rm Tr}$} & \multicolumn{2}{c|}{MLP} & \multicolumn{2}{c|}{CNN}  & \multicolumn{2}{c}{GNN} \\ \cline{3-8}  &    & \multicolumn{1}{c|}{OP}   & FR    & \multicolumn{1}{c|}{OP}      & FR   & \multicolumn{1}{c|}{OP}   & FR   \\
\hline
    MMSE & 30  & \multicolumn{1}{c|}{94.09\%} & 100\% & \multicolumn{1}{c|}{95.12\%} & 100\% & \multicolumn{1}{c|}{95.14\%} & 100\%  \\
\hline
    HZM & 30   & \multicolumn{1}{c|}{94.56\%} & 100\% & \multicolumn{1}{c|}{94.62\%} & 100\% & \multicolumn{1}{c|}{98.48\%} & 100\%  \\
\hline 
\hline
    \multirow{2}{*}{Method} & \multirow{2}{*}{$K_{\rm Te}$} & \multicolumn{2}{c|}{MLP}  & \multicolumn{2}{c|}{CNN}  & \multicolumn{2}{c}{GNN}  \\ 
    \cline{3-8} 
     &   & \multicolumn{1}{c|}{SP} & FR & \multicolumn{1}{c|}{SP} & FR & \multicolumn{1}{c|}{SP}  & FR \\ 
\hline
    \multirow{6}{*}{MMSE}   & 15                 & \multicolumn{1}{c|}{85.54\%} & 100\% & \multicolumn{1}{c|}{86.57\%} & 100\% & \multicolumn{1}{c|}{85.62\%} & 100\%  \\ 
    \cline{2-8} 
    & 20   & \multicolumn{1}{c|}{89.09\%} & 100\% & \multicolumn{1}{c|}{89.98\%} & 100\% & \multicolumn{1}{c|}{89.06\%} & 100\%  \\ \cline{2-8} 
    & 35  & \multicolumn{1}{c|}{×}       & × & \multicolumn{1}{c|}{×} & × & \multicolumn{1}{c|}{96.98\%} & 100\%  \\ 
    \cline{2-8} 
    & 40 & \multicolumn{1}{c|}{×}       & × & \multicolumn{1}{c|}{×}   & ×  & \multicolumn{1}{c|}{98.44\%} & 98.2\% \\ 
    \cline{2-8} 
    & 45  & \multicolumn{1}{c|}{×} & × & \multicolumn{1}{c|}{×}  & × & \multicolumn{1}{c|}{98.53\%} & 75.3\% \\ 
    \cline{2-8} 
    & 50  & \multicolumn{1}{c|}{×}       & × & \multicolumn{1}{c|}{×}       & ×  & \multicolumn{1}{c|}{0\%} & 0\% \\ 
\hline
    \multirow{6}{*}{HZM}  & 15     & \multicolumn{1}{c|}{86.05\%} & 100\% & \multicolumn{1}{c|}{86.46\%} & 100\% & \multicolumn{1}{c|}{86.28\%} & 100\%  \\ 
    \cline{2-8} 
    & 20  & \multicolumn{1}{c|}{89.92\%} & 100\% & \multicolumn{1}{c|}{89.82\%} & 100\% & \multicolumn{1}{c|}{90.53\%} & 100\%  \\ \cline{2-8} 
    & 35 & \multicolumn{1}{c|}{×}       & × & \multicolumn{1}{c|}{×} & × & \multicolumn{1}{c|}{97.59\%} & 100\%  \\ 
    \cline{2-8} 
    & 40  & \multicolumn{1}{c|}{×}    & ×  & \multicolumn{1}{c|}{×}    & ×  & \multicolumn{1}{c|}{93.65\%} & 100\%  \\ 
    \cline{2-8} 
    & 45  & \multicolumn{1}{c|}{×}       & ×  & \multicolumn{1}{c|}{×}  & × & \multicolumn{1}{c|}{89.94\%} & 100\%  \\ 
    \cline{2-8} 
    & 50  & \multicolumn{1}{c|}{×}   & ×& \multicolumn{1}{c|}{×}   & ×  & \multicolumn{1}{c|}{86.18\%} & 57.6\% \\ 
\hline
\end{tabular}\label{user scale}
     \begin{tablenotes}
        \footnotesize
        \item{$K_{\rm Te}$: Value of $K$ in the test set.}
        \item {OP/SP/FR: Optimality performance/scalability performance/feasibility rate.}
\end{tablenotes}
\end{table*}

Table \ref{environmental parameter} shows the scalability performance to noise power and rate requirement of the GNN. The feasibility rate for all simulated cases is $100\%$. The GNN trained with $(\Gamma,{\xi})=(0.5,1)$ (Dataset No. 11) are test with $15$ different $(\Gamma,{\xi})$ settings (Dataset No. 3, 4, 5, 6, 7, 8, 9, 10, 11, 12, 13, 14, 15, 16, 17). It is observed that the model-based GNN still maintains a satisfying performance when facing unseen channel conditions and QoS requirements during the training phase. Similar to Table \ref{base experiment}, the MMSE scheme is with better scalability performance in the low-SNR and high-SNR regions while the HZM scheme is more preferable for other cases. Thanks to the scheme selection module, the scalability performance of the model-based GNN is further enhanced by integrating the advantages of both MMSE and HZM schemes. Nevertheless, it should note that the feasibility rate is poor for both MMSE and HZM schemes when the GNN scales to the scenario with $K=50$. As the feasibility rate stands for the QoS performance, it is crucial to enhance the feasibility rate. To this end, we compare the constant-input and various-input training strategies as follows.

\begin{table*}[t]
    \centering
        \caption{Scalability performance to number of MUs: ${\Gamma_{\rm Tr}=0.5}$ and ${{\xi}_{\rm Tr}=1}$}
\begin{tabular}{c|c|c|c||c|c|c}
\hline
Method &  \multicolumn{3}{c||}{MMSE} & \multicolumn{3}{c}{HZM} \\
\hline
\diagbox{${\Gamma_{\rm Te}}$}{${\xi}_{\rm Te}$} & \multicolumn{1}{c|}{0.5} & \multicolumn{1}{c|}{0.75} & \multicolumn{1}{c||}{1} & \multicolumn{1}{c|}{0.5} & \multicolumn{1}{c|}{0.75} & \multicolumn{1}{c}{1}\\ \hline
0.1     &   \textbf{81.67\%} & \textbf{80.87\%} &  \textbf{81.63\%} &  77.46\% & 77.35\% &   77.37\%  \\ \hline
0.25    &   89.29\% & 89.74\% & 89.68\% & \textbf{90.03\%} & \textbf{ 90.03\%} &  \textbf{90.02\%} \\ \hline
0.5 & 95.16\% &  94.11\% &  94.22\%$^*$ & \textbf{98.43\%} & \textbf{98.35\%} &  \textbf{98.48\%}$^*$    \\ \hline
0.75    & 95.75\% & 96.61\% & 96.63\% &  \textbf{97.70\%} &  \textbf{97.56\%} & \textbf{97.54\%}  \\ \hline
1       & \textbf{98.90\%} & \textbf{97.96\%} & \textbf{98.86\%} & 95.76\% &  95.58\% &  95.53\% \\ \hline
\end{tabular}
\label{environmental parameter}
     \begin{tablenotes}
        \footnotesize
        \item{$^*$: The result marked with $^*$ represents the optimality performance.}
\end{tablenotes}
\end{table*}


\subsection{Evaluation of Training Strategy}

Table  \ref{training strategy} evaluate the  scalability performance to number of MUs and feasibility rate of the model-based GNN via constant-input (Dataset No. 11) and various-input (Dataset No. 22) training strategies. The scheme selection module is activated. Note that the numbers of the training samples are equivalent for the two training strategies. It is observed that the various-input training strategy achieves similar scalability performance with the constant-input strategy in general while the former one greatly enhances the feasibility rate for $K\in\{40,45,50\}$. The reason is that the various-input training strategy allows the GNN to learn more problem sizes. Besides, the different training strategies may result in different schemes. Moreover, for $K=50$, the feasibility rate is enhanced by the various-input training strategy at expense of the degradation of the scalability performance. Therefore, a trade-off may exist between the scalability performance and the feasibility rate for GNN.

\begin{table}[t]
    \centering
    \caption{Performance evaluation of training strategies}
    \begin{tabular}{c|c|c|c|c}
    \hline
         {${K_{\rm Tr}}$} &{$K_{\rm Te}$} &  {OP} &{FR}  &{SS} \\
    \hline
    \hline
         
         & 15  & 86.28\% & 100\% & HZM\\
         \cline{2-5}
         & 20  &90.52\% & 100\% &  HZM \\
         \cline{2-5}
         & 30  &98.48\%$^\dag$ & 100\%$^\dag$ & HZM\\
         \cline{2-5}
        30 & 35  &97.59\% & 100\% & HZM\\
         \cline{2-5}
        (constant-input) & 40  & 98.44\% & 98.2\% & MMSE\\
         \cline{2-5}
         & 45  & 98.53\% & 75.3\% & MMSE\\
         \cline{2-5}
         & 50  & 85.69\% & 57.6\% & HZM\\
    \hline
         & 15  & 86.47\% & 100\% & MMSE\\
         \cline{2-5}
         & 20  & 89.89\% & 100\% & MMSE\\
         \cline{2-5}
         & 30  & 95.01\%$^\dag$ & 100\%$^\dag$ & HZM\\
         \cline{2-5}
       $\{30,40,50\}$  & 35  & 99.20\% & 100\% & HZM\\
         \cline{2-5}
        (various-input) & 40  & 98.31\%$^\dag$ & 99.95\%$^\dag$ & MMSE\\
         \cline{2-5}
         & 45  & 99.20\% & 91.05\% & MMSE\\
         \cline{2-5}
         & 50  & 80.18\%$^\dag$ & 94.15\%$^\dag$ & HZM\\
    \hline
    \end{tabular}\label{training strategy}
         \begin{tablenotes}
        \footnotesize
        \item{SS: selected scheme.}
        \item{$^\dag$: The result marked with $^\dag$ represents that $K_{\rm Te}$ is seen in the training set.}
\end{tablenotes}
  
\end{table}

\section{Conclusion}

In this paper, we studied a model-based GNN enabled beamforming design for a QoS constrained EE maximization problem in ultra-dense wireless networks. To increase the learning performance and computational efficiency, the prior knowledge based on MMSE and HZM schemes was utilized to reformulate the considered problem. To enhance the feature extracting, the attention mechanisms and residual structures were applied. To improve the stability of the model-based GNN when applying in unseen scenarios, the various-input training strategy was proposed. The model-based GNN was trained via unsupervised learning and evaluated from the perspectives of optimality, scalability, feasibility rate and inference time. Compared with the existing CVX, MLP and CNN, the advantages of the model-based GNN in ultra-dense wireless networks are summarized as: 1) real-time response to time-varying CSI with (near-)optimal performance; 2) good scalability to MUs; 3) good adaptability to various channel conditions and QoS requirement.

\appendix

\subsection{Single-loop solution approach for the problem $\rm P_0$}

\label{slsa}

Most existing algorithms for energy-efficient beamforming design are based the Dinkelbach method and semidefinite relaxation (SDR). This paper proposes a new algorithm without Dinkelbach method and SDR for solving the considered problem $\rm P_0$ to improve the computational efficiency and stability.

Define the following auxiliary variable to replace the objective function of the problem $\rm P_{0}$.
\begin{flalign}\label{ee}
{\varpi _k} \triangleq \frac{{ {{R_k}\left( {\left\{ {{{\bf{w}}_i}} \right\}} \right)} }}{P_{\rm Total}\left( \{{{\bf{w}}_i}\} \right)}
\end{flalign}
Then, solving the problem $\rm P_{0}$ is equivalent to maximizing $\sum\nolimits_{k\in {\cal K}} {{\varpi _k}}$ subject to the constraints of the problem $\rm P_{0}$ and 
\begin{flalign}
{R_k}\left( \{{{\bf{w}}_i}\} \right) \ge {\varpi _k}{P_{\rm Total}\left( \{{{\bf{w}}_i}\} \right)}.\label{PC:1}
\end{flalign}
That is, one can reformulate the problem $\rm P_{B2}$ as the following problem:
\begin{subequations}\label{p2}
\begin{align}
{\bf P_2:}~&\mathop {\max }\limits_{\left\{{{{\bf{w}}_i}},{\varpi _i}\ge0 \right\}}\sum\nolimits_{k\in {\cal K}} {{\varpi _k}} \\
{\rm s.t.}~&\eqref{cons:A},~\eqref{cons:B},~\eqref{cons:C},~\eqref{PC:1}.
\end{align}
\end{subequations}

By defining
\begin{flalign}\label{slack_v2}
\left\{ \begin{array}{l}
{e^{{a_{k}}}}\triangleq{\varpi _k},\\
{e^{{b}}}\triangleq{{\sum\nolimits_{i\in {\cal K}} {{{\left\| {{{\bf{w}}_i}} \right\|}^2}}  + {P_{\rm{C}}}}},\\
c_k \triangleq {e^{{a_{k}}}} + {e^{{b}}},\\
e^{d_k} \triangleq {{2^{{c _k}}} - 1},\\
e^{f_k} \triangleq {\sum\nolimits_{i \in {{\cal K}}/\{ k\} } {{{\left| {{\bf{h}}_k^H{{\bf{w}}_i}} \right|}^2}}  + \sigma _k^2},
\end{array} \right.
\end{flalign} and substituting (\ref{slack_v2}) into  the problem $\rm P_2$, one can reformulate the problem $\rm P_{2}$ as the following problem:
\begin{subequations}\label{p2a}
\begin{align}
&{\bf P_{2-A}:}~\mathop {\max }\limits_{\left\{{{{\bf{w}}_i}},{\varpi _i}\ge0,a_i,b,c_i,d_i,f_i\in {\mathbb R}  \right\}}\sum\nolimits_{k\in {\cal K}} {{\varpi _k}} \\
{\rm s.t.}~&
{\left| {{\bf{h}}_k^H{{\bf{w}}_k}} \right|^2} \ge e^{\left(d_k + f_k\right)},
\label{p2a:a}\\
&e^{d_k} \ge {{2^{{c _k}}} - 1},\label{p2a:b}\\
&e^{f_k} \ge {\sum\nolimits_{i \in {{\cal K}}/\{ k\} } {{{\left| {{\bf{h}}_k^H{{\bf{w}}_i}} \right|}^2}}  + \sigma _k^2},\label{p2a:c}\\
&{c_k} \ge {e^{\left( {{a_k} + b} \right)}},\label{p2a:d}\\
&e^{a_k}\ge{\varpi _k},\label{p2a:e}\\
&e^{b}\ge{{\sum\nolimits_{i\in {\cal K}} {{{\left\| {{{\bf{w}}_i}} \right\|}^2}}  + {P_{\rm{C}}}}},\label{p2a:f}\\
&{\left| {{\bf{h}}_k^H{{\bf{w}}_k}} \right|^2} \ge \left( {{2^{{\xi _k}}} - 1} \right)\left( {\sum\limits_{i \in {{\cal K}}/\{ k\} } {{{\left| {{\bf{h}}_k^H{{\bf{w}}_i}} \right|}^2}}  + \sigma _k^2} \right), \label{p2a:g}\\
&{\rm \eqref{cons:B},~\eqref{cons:C}.}\nonumber
\end{align}
\end{subequations}
With a given feasible point $\{{{\widetilde a}_{i}},{{\widetilde{ b}}},{{\widetilde d}_{i}},{{\widetilde f}_{i}},{{{\widetilde {\bf{ w}}}_i}}\}$, applying their first-order Taylor approximations gives rise to the following approximated convex problem:
\begin{subequations}
\begin{align}
&{\bf P_{2-B}:}~\mathop {\max }\limits_{\left\{{{{\bf{w}}_i}},{\varpi _i}\ge0,a_i,b,c_i,d_i,f_i\in {\mathbb R}  \right\}}\sum\nolimits_{k\in {\cal K}} {{\varpi _k}} \\
{\rm s.t.}~& {2{{\rm Re}} \left\{ {\widetilde {\bf{ w}}_k^H{{\bf{h}}_k}{\bf{h}}_k^H{{\bf{w}}_k}} \right\} - {\left| {{\bf{h}}_k^H{{{{\widetilde {\bf{ w}}}_k}}}} \right|^2}} \ge e^{\left(d_k + f_k\right)},
\label{p2b:a}\\
&{e^{{{\widetilde d}_k}}}\left( {{d_k} - {{\widetilde d}_k} + 1} \right) \ge {{2^{{c _k}}} - 1},\label{p2b:b}\\
&{e^{{{\widetilde f}_k}}}\left( {{f_k} - {{\widetilde f}_k} + 1} \right) \ge {\sum\limits_{i \in {{\cal K}}/\{ k\} } {{{\left| {{\bf{h}}_k^H{{\bf{w}}_i}} \right|}^2}}  + \sigma _k^2}\label{p2b:c}\\
&{e^{{{\widetilde a}_k}}}\left( {{a_k} - {{\widetilde a}_k} + 1} \right)\ge{\varpi _k},\label{p2b:d}\\
&{e^{{{\widetilde b}}}}\left( {{b} - {{\widetilde b}} + 1} \right)\ge{{\sum\nolimits_{i\in {\cal K}} {{{\left\| {{{\bf{w}}_i}} \right\|}^2}}  + {P_{\rm{C}}}}},\label{p2b:e}\\
& {2{{\rm Re}} \left\{ {\widetilde {\bf{ w}}_k^H{{\bf{h}}_k}{\bf{h}}_k^H{{\bf{w}}_k}} \right\} - {\left| {{\bf{h}}_k^H{{{{\widetilde {\bf{ w}}}_k}}}} \right|^2}} \ge \label{p2b:f}\\
&\left( {{2^{{\xi _k}}} - 1} \right)\left( {\sum\nolimits_{i \in {{\cal K}}/\{ k\} } {{{\left| {{\bf{h}}_k^H{{\bf{w}}_i}} \right|}^2}}  + \sigma _k^2} \right), \nonumber \\
&{\rm \eqref{cons:B},~\eqref{cons:C},~\eqref{p2a:d}.}\nonumber
\end{align}
\end{subequations}

To improve the approximation precision of the first-order Taylor approximation, the SCA method can be  employed to iteratively solve the problem $\rm P_{2-B}$.

\end{document}